\documentclass[twocolumn,superscriptaddress]{revtex4}

\usepackage{amsmath}  
\usepackage{amsfonts} 
\usepackage{graphicx} 
\usepackage{color}
\usepackage{siunitx}
\usepackage{subfig}
\usepackage{blindtext}

\begin{document}


\title{Weighing the Sun with five photographs}

\author{Hugo Caerols}
\email{hugo.caerols@uai.cl}
\affiliation{Facultad de Ingenier\'{\i}a y Ciencias, Universidad Adolfo Ib\'a\~nez, Santiago 7941169, Chile.}
\affiliation{Grupo de Observaci\'on Astron\'omica OAUAI, Universidad Adolfo Ib\'a\~nez, Santiago 7941169, Chile. }

\author{Felipe A. Asenjo}
\email{felipe.asenjo@uai.cl}
\affiliation{Facultad de Ingenier\'{\i}a y Ciencias, Universidad Adolfo Ib\'a\~nez, Santiago 7941169, Chile.}


\date{\today}

\begin{abstract}
With only five photographs of the Sun at different dates we  show that the mass of Sun can be calculated by using a telescope, a camera, and the  Kepler's third law. With these photographs we are able to calculate the distance between   Sun and Earth at different dates in a period of time of about three months. These distances allow us to obtain the correct elliptical orbit of Earth, proving the  Kepler's first law.
 The analysis of the data extracted from photographs
is performed by using an analytical optimization approach that allow us to find the parameters of the elliptical orbit. Also, it is shown that the five data points fit an ellipse
using an  geometrical scheme.  The obtained parameters  are in very good agreement with the ones for Earth's orbit, allowing us to foresee the future positions of Earth along its trajectory. 
The parameters for the orbit are used to calculate the Sun's mass by applying the  Kepler's third law and Newton's law for gravitation.
This method gives a result wich is in excellent agreement with the correct value for the Sun's mass. Thus, in a span of time of about three months, any student is capable to calculate the mass of the sun with only five photographs, a telescope and a camera.
\end{abstract}
 
\maketitle 

\section{Introduction}

Our solar system is completly dominated by the Sun. Its huge mass bounds the eight planets in elliptical orbits around it. The proper explanation of the motion of each planet can be found from the Newton's  law of universal gravitation. One of the greatest triumph of Newton's law is to provide the physics foundation to the empirical observations encompassed in the  Kepler's three laws. 

Understanding  Kepler's laws allow us to calculate the relation between the shape of the orbit of each planet around Sun and the amount of time (the period) that takes one revolution around it. This relation depends only on the mass of the Sun in our solar system, and usually is used to calculate the distances to it or the period of the celestial object orbiting it. However, it can be used in the opposite way. If the  perihelion and aphelion distances, and the period are known, then the mass of the central star can be calculated.

The purpose of this work is to draw attention that the mass of the Sun can be accurately calculated with only five photographs of the Sun taken from Earth. Our aim is to show that any student can perform this project in an very easy way in an span of time of about three months.  Only five photographs are needed as they provide five data points for the Earth's position around the Sun. With those five points, always a conic curve can be found to fit them. As we will show below, that conic curve will correspond to an ellipse. Thus, knowing the ellipse properties, we are able to calculate the Sun's mass.

All this can be done by using Kepler's third law for planetary motion.
 This law relates the parameters of the elliptical orbit of a planet with the mass of the central star in which it revolves. 
{When the star is at one of the focus of the elliptical orbit, the third law tell us that the cube of the semimajor axis $a$ of the orbit is proportional to the square of the period $T$ of one revolution around that star, i.e., the ratio $a^3/T^2$ is a constant. Finding the value of that constant was one of the greatest achievements of Newton's law of gravitation.
Mathematically, the  Kepler's third law can be derived from first principles from Newton's law of gravitation \cite{thomasc}, to simply be written as
\begin{equation}\label{secondKlaw}
\frac{a^3}{T^2}=\frac{GM}{4\pi^2}\, ,
\end{equation}
where $G=6.6726\times 10^{-11}$[$\mbox{m}^3\mbox{s}^{-2}/\mbox{Kg}$] is the universal gravitational constant \cite{nrl}, and $M$ is the mass of the central star. 

It is the calculation of  parameters $a$ and $T$} (and their combination) that allow us to estimate the mass of the Sun. Some of the parameters are need to be measured (the purpose of this work), while others are assumed to be known. For our proposal, we consider known quantities as
 the universal gravitational constant, the period of time of Earth (the time that takes to perform one revolution), and the diameter of Sun. With these known physical quantities, we are able to estimate the semimajor distance of Earth's orbit, its eccentricity and the Sun's mass, both from five photographs.

According to   NASA \cite{nasasunshhet}, the values for the Sun's mass and diameter are 
\begin{eqnarray}
M_\odot&=&1.9885\times 10^{30}\, \mbox{[Kg]} \, ,\label{datossolnasaMASA}\\
D_\odot&=&1.3914\times 10^6\, \mbox{[Km]}\, .\label{datossolnasa}
\end{eqnarray}
Also, Earth's orbit is elliptical, with an orbit eccentricity $\epsilon_E$  and a semimajor axis $a_E$ given by \cite{nasasunshhet2}
\begin{eqnarray}
\epsilon_E&=&0.0167 \, ,\label{datossolnasaEx2}\\
a_E&=&1.496\times 10^8\, \mbox{[Km]}\, .\label{datossolnasasemiejer}
\end{eqnarray}
Below, we show how with five photgraphs, the values of parameters \eqref{datossolnasaEx2} and \eqref{datossolnasasemiejer}
can be accuratelly determined. 
 In order to perform the estimation for the Sun's mass, only the following few assumptions are required:
\begin{enumerate}
\item Earth revolves around Sun.
\item The time that Earth takes to perform one revolution around the Sun (a year) is  known, given by
\begin{equation}\label{period}
T=365.25\, [\mbox{days}]=3.1558\times 10^7\,  [\mbox{s}]\, .
\end{equation} 
\item It is assumed that the Sun's diameter \eqref{datossolnasa} is known.
\end{enumerate}
The above reasonable  assumptions are the key to estimate the Sun's mass. The photographs taken to the Sun are used to calculated the distance from Earth to it at five different dates. These distances and dates correspond to the position of Earth with respect to Sun, allowing us to calculate the Sun's mass using Kepler's laws.
In the precceding sections, we will show how from the five photographs we are able to estimate the Sun's mass \eqref{datossolnasaMASA} with a percentage error of about 0.1\%.

The most sensitive part of this project is the photograph analysis, which is thoroughly explained in Sec.~\ref{photsunsection}. It is from it  that  distances from Sun to Earth can be obtained by relating the telescope angles (telescope features) and the physical angles and distances between the two astronomical objects. Thus, by measuring angles in the images, we are able to measure different distances.
 Then, in Sec.~\ref{polarajuste}, the  approach to use the data obtained from the photographs is explained, showing that it allows us to find an ellipse with parameters  in agreement with \eqref{datossolnasaEx2} and \eqref{datossolnasasemiejer}.
Besides, in Sec.~\ref{Keplersections} we use the results from previous sections to calculate the Sun's mass, and to discuss how our data adjust to the  Kepler's three laws.
Finally, in the last section, we comment the validity of our assumptions.

\section{Photographs of the Sun and its distance estimation}
\label{photsunsection}

\begin{figure*}[ht]
\begin{tabular}{cc}
  \includegraphics[width=70mm]{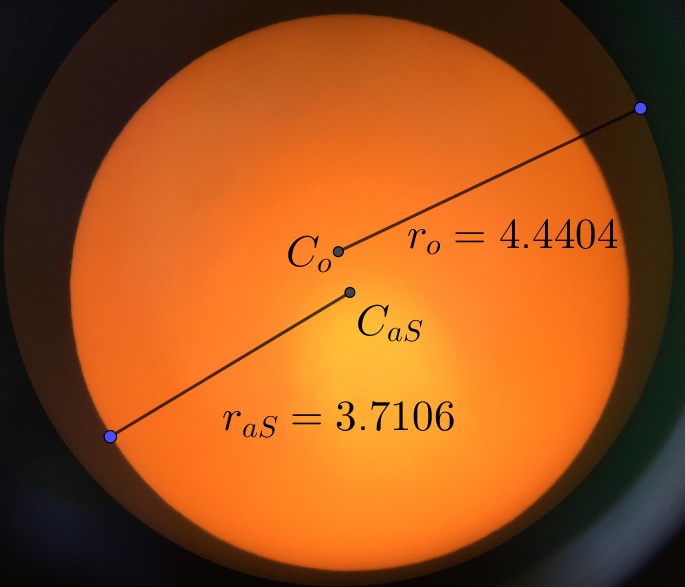} &   \includegraphics[width=71mm]{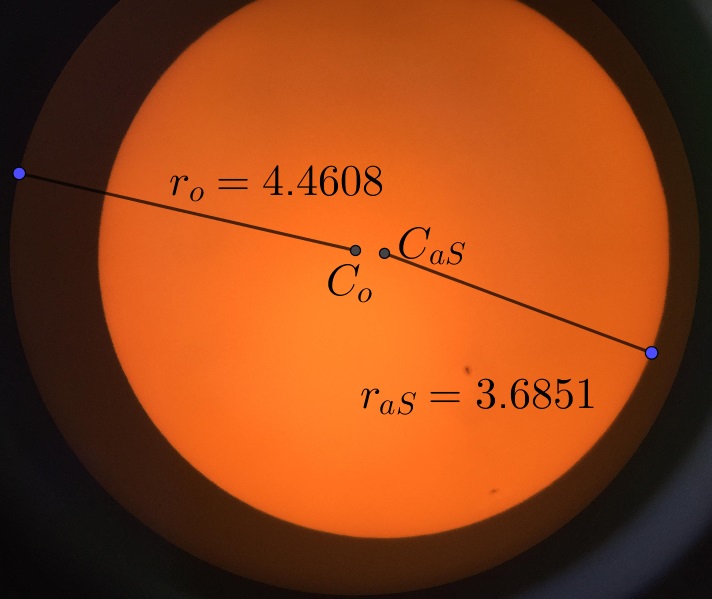} \\
(a) March 26th, 2019 & (b)  May 7th, 2019 \\[6pt]
 \includegraphics[width=61mm]{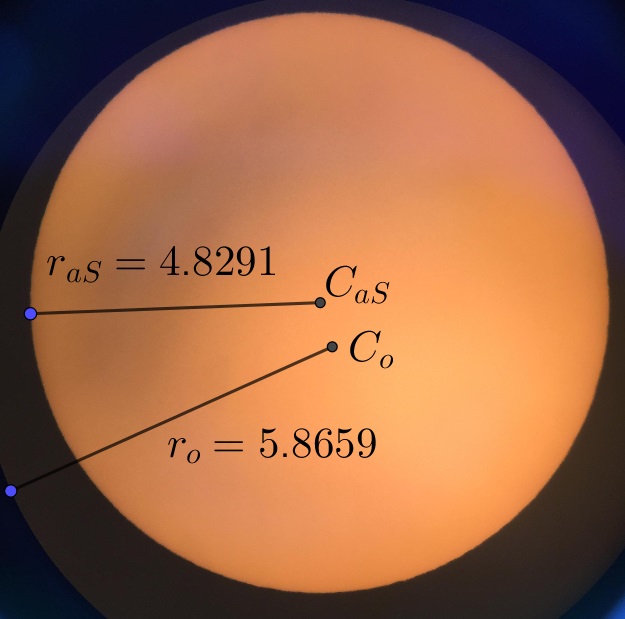} &   \includegraphics[width=70mm]{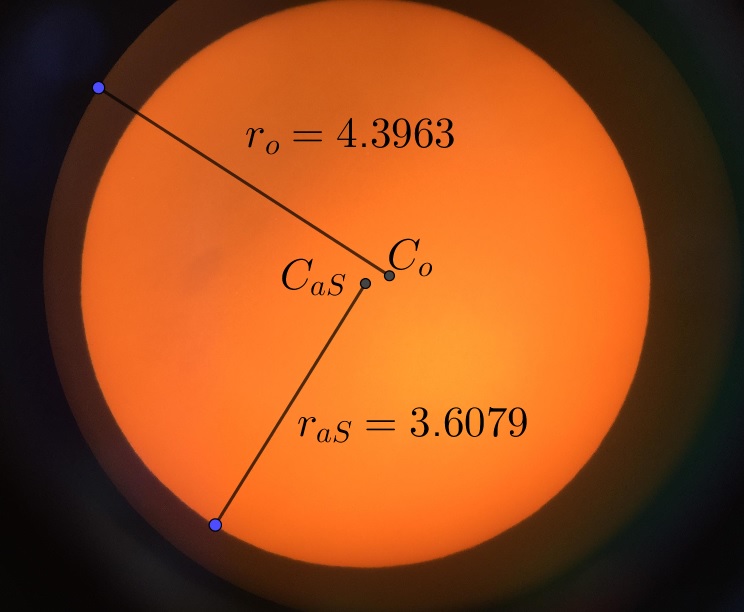} \\
(c) May 20th, 2019 & (d) June 8th, 2019 \\[6pt]
\multicolumn{2}{c}{\includegraphics[width=70mm]{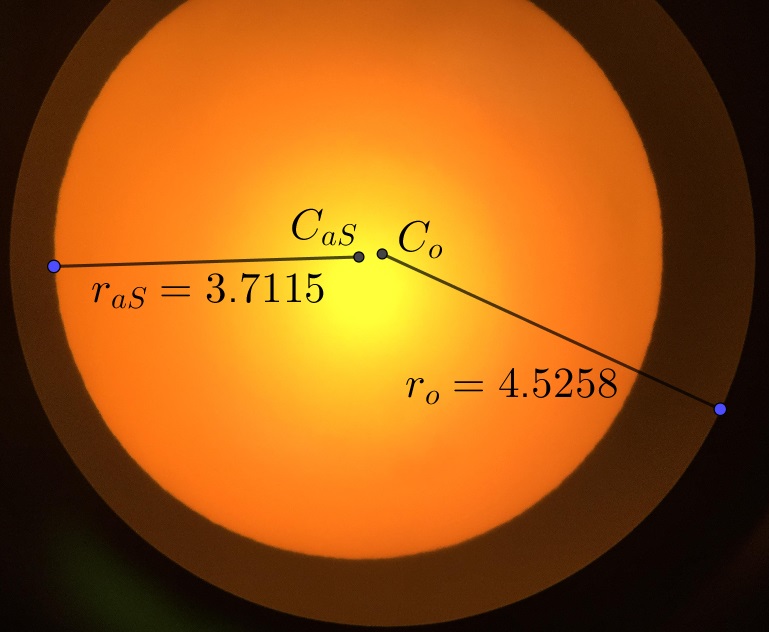} }\\
\multicolumn{2}{c}{(e) June 21th, 2019}
\end{tabular}
\caption{Five photographs of the Sun at different dates. Here, $r_o$ is the  effective radius of the ocular, and $r_{aS}$ is the effective radius of the image (Sun).}
\label{eclipseFig}
\end{figure*}

During 2019, from March to June, five photographs of the Sun were taken from Santiago, Chile. The photographs are shown in Fig.~\ref{eclipseFig}.
The used telescope for this experience is a Celestron Nexstar 8SE. Its focal length is $2032$[mm], and its magnification is \ang{52} with the ocular  focal length of $25$[mm].  It was used a Eclipsmart Solar Filter 8” SCT attached to the telescope. Also, an Iphone 6 was used to take the photographs attached to the telescope with
a NexYZ 3-Axis Universal Smartphone adapter.
The photographs, taken in several different dates, were used to calculate the distance from Earth to  Sun. The only requirement is that the image of the Sun  is completely inside of the ocular of the telescope. Once that condition is achieved, the method to calculate the distance is straightforward.
\begin{figure}[h!]
\centering
\includegraphics[width=6cm]{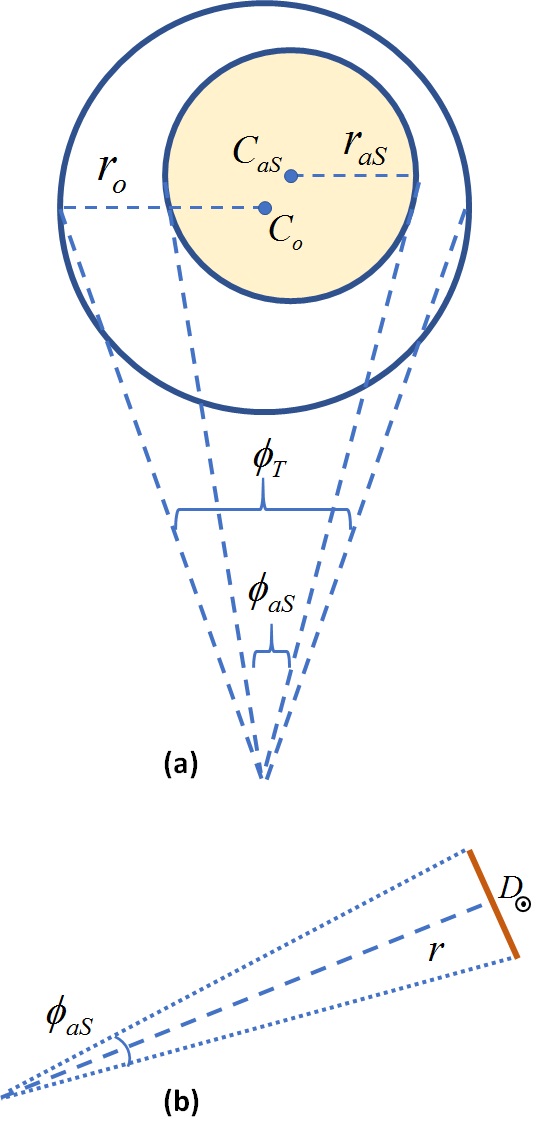}  
\caption{(a) Scheme of the relation between the apparent ocular angle of an image $\phi_{aS}$ and the  effective ocular angle $\phi_T$ of the telescope. (b) Relation between the  the apparent ocular angle of an image $\phi_{aS}$, its physical size $D_\odot$ and  its distance $r$ to the observer.}
\label{figuratelescopesuncal}
\end{figure}

By using the features of the telescope, and compare them with the images obtained, the physical distances to the observed objects can be inferred.  The procedure is explained below, and it is depicted in Fig.~\ref{figuratelescopesuncal}.
A given telescope has an ocular focal length $d_o$, a focal length $f$, and an apparent field of view $m$. The telescope magnification is determined as $f/d_o$, and the effective ocular angle  of the telescope $\phi_T$ can be obtained as
\begin{equation}
\phi_T=\left(\frac{\pi}{180}\right)\left(\frac{m\,  d_o}{f}\right)\, ,
\end{equation} 
measured in radians. For the telescope used in this work, $d_o=25$[mm], $f=2032$[mm], and $m=$ \ang{52}, giving $\phi_T\approx 0.011166$.

\begin{figure}[h!]
\centering
\includegraphics[width=7.5cm]{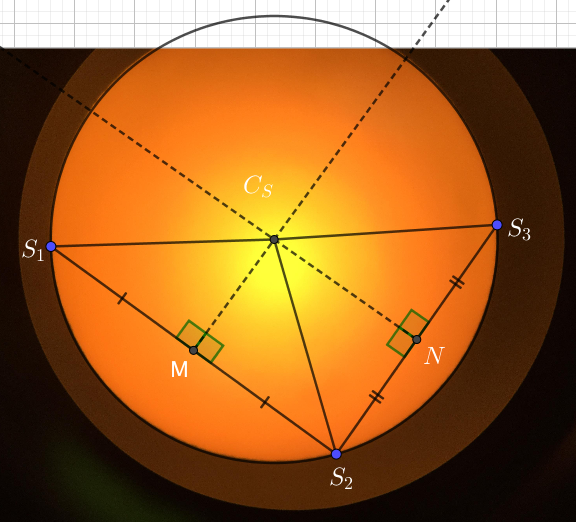}  
\caption{(a) Method to calculate the center and  radius of Sun in an image using GeoGebra. For the ocular, the same procedure applies.}
\label{circuncenterSUN}
\end{figure}

Once $\phi_T$ is found, it can be used  to obtain the aparent ocular angle $\phi_{aS}$ of any object  that image is  lying completely inside the ocular size of the telescope, as it is shown in Fig.~\ref{figuratelescopesuncal}(a).  Let us assume that for a given photograph (such as those shown in Fig.~\ref{eclipseFig}) we calculate the effective radius $r_o$ of the ocular and the effective radius $r_{aS}$ of the image (of the Sun in our cases). As it is shown in Fig.~\ref{eclipseFig}, those radii can be calculated with GeoGebra software \cite{geogebra} using its own scale. However, this is not relevant as it is only important  their relative magnitude.
This geometry free-source  software allows to select points in the image, which are used to calculate the different radii. These radii can then obatined using geometry, and the general method is presented in Fig.~\ref{circuncenterSUN}. In general, only three points are need to determine the radius of the circumscribed circumference \cite{cox}. Consider a circumference with undetermined radius. Let us choose three points  in its perimeter, as $S_1$, $S_2$ and $S_3$. The segments  $\overline{S_1 S_2}$ and $\overline{S_2 S_3}$ connect these three points. Now, GeoGebra alow us to calculate the mediatrix 
to each segment  at points $M$ and $N$. The key step is to notice that mediatrixes intersect at $C_S$. In this form, we can  construct two isosceles triangles, that  share one side. Therefore, the intersection point is the center of the circle, 
the sides represent the radius of the circumference circumscribed to the three initial points, and  Geogebra measures the segments in units of the software. We do this for the image of Sun and the ocular.

Once the two radii are known, then the apparent ocular angle of the  image (of the Sun) is given simply by
\begin{equation}
\phi_{aS}=\left(\frac{r_{aS}}{r_o}\right)\phi_T\, .
\label{eqdistanSun0}
\end{equation}
The apparent ocular angle $\phi_{aS}$ is the most important parameter to be determined in an image,
as it allow us to relate the real physical size of the image with its physical distance $r$ to the observer that took the photograph. As it can been seen in Fig.~\ref{figuratelescopesuncal}(b), if the physical size of the image (in this case the diameter  $D_\odot$ of the Sun) is known, then it can be modelled as the arc sustended by the ocular angle $\phi_{aS}$ at distance $r$, such that $D_\odot=r \, \phi_{aS}$. This is possible as  the scales of the Sun are smaller than its distance from Earth. Therefore, we find that the distance from Earth to the Sun is
\begin{equation}
r=\frac{D_\odot}{\phi_{aS}}=D_\odot \left(\frac{r_{o}}{r_{aS}}\right)\left(\frac{180 f}{\pi\,  m\, d_o} \right)\, .
\label{eqdistanSun}
\end{equation}

In our case, when the diameter of the Sun is assumed  with a value given in Eq.~\eqref{datossolnasa}, we can readily calculate  distances $r$  by using  photographs and Eq.~\eqref{eqdistanSun}. 
In Table \ref{bosons}, we tabulate the different photographs (ordered by date) with their corresponding values of $r_o$ and  $r_{aS}$ measured directly from the photographs (see Fig.~\ref{eclipseFig}), together with their calculated values of $\phi_{aS}$ and $r$ obtained from Eqs.~\eqref{eqdistanSun0} and \eqref{eqdistanSun}. The radii $r_o$ and $r_{aS}$ depicted in Table \ref{bosons} are measured in centimeters units using the GeoGebra software. The measurement units of these quantities is not important, as we only require the information of its ratio.  The radii $r_o$ and $r_{aS}$ have four significant figures due to the camera of Iphone 6 has photographs with 3264$\times$2448 pixels. Thus, the distance $r$ in Eq.~\eqref{eqdistanSun} have only four significant figures. 
\begin{table*}[htp]
\begin{ruledtabular}
\begin{tabular}{c c c c c c}
photo & date  & $r_o\, \mbox{(in cm)}$ & $r_{aS}\, \mbox{(in cm)}$ & $\phi_{aS}$  &  $r\, \mbox{(in Km)}$  \\
\hline	
1& March 26th, 2019 & 4.4404 &  3.7106  & 0.009331 & 149118984.1571\\
2& May 7th, 2019   & 4.4608 &  3.6851 & 0.009224 & 150840671.5767\\
3& May 20th, 2019   & 5.8659  & 4.8291   & 0.009192 & 151364293.7829 \\
4& June 8th, 2019   & 4.3963 &  3.6079 & 0.009164 & 151840565.2067  \\
5& June 21th, 2019  &  4.5258 & 3.7115 & 0.009157 & 151950060.0148
\end{tabular}
\end{ruledtabular}
\centering
\caption{Estimation of distances $r$ to Sun, given in Eq.~\eqref{eqdistanSun}, at different dates by using telescope and images measurements.}
\label{bosons}
\end{table*}

\begin{figure}[h!]
\centering
\includegraphics[width=6.5cm]{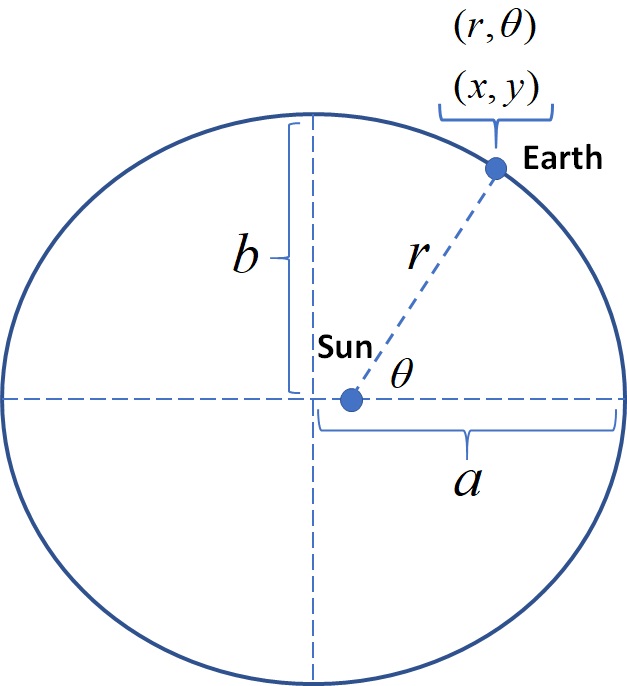}  
\caption{Schematic drawing of the Earth's elliptical orbit around the Sun (at one of the focus). Earth's position depends on the distance to the Sun and on its date along the year. This can be mathematically represented in polar coordinates $(r,\theta)$, or in cartesian coordinates $(x,y)$. At the perihelion, $\theta=0$.}
\label{coordinatesfigure}
\end{figure}

At different dates, the Sun is at different distances $r$ from Earth. As we know that Earth is in orbit around the Sun, then the data shows that the orbit is not circular.
In order to calculate the shape of the Earth's orbit we need to know the position of Earth during the whole timespan of the data acquisition. As the orbital trajectory is a two-dimensional curve, we need two coordinates to describe the position of Earth. As it is shown in Fig.~\ref{coordinatesfigure}, the coordinates can be defined using the distance from Sun to Earth and its angular position (called polar coordinates), or using the Earth position with respect to the Sun in a cartesian plane (called cartesian coordinates). Both systems are explored in the following sections. 

From Table \ref{bosons}, we already have the magnitude of the distance $r$ for different dates, but we lack the information on the angular position $\theta$ of Earth along the trajectory (see Fig.~\ref{coordinatesfigure}). This angle can be defined from the perihelion (the closest distance from Earth to Sun) by choosing it when  $\theta=0$.
We can  estimate the relation between date and the angular position of Earth in a very simple manner. First, as we known that  Earth is revolving in a closed orbit, then let us assume that $\theta=2\pi$ in one year. Second,  let us assume that Earth revolves around Sun at a constant rate. With these considerations, the angular position $\theta$ can be related with the date of each photograph in the simplest possible way as
\begin{equation}\label{thetaconstantrate}
\theta=\frac{2\pi\, t}{T}\, ,
\end{equation}
measured in radians. Here,  $T$ is the time that last one terrestial year given in days by Eq.~\eqref{period}, and
$t$ is the time (in the same units than $T$) that have passed from the perihelion of the orbit.
For Earth, during 2019, the perihelion occurred approximatelly in the night of January, 2nd.
The relation \eqref{thetaconstantrate} between the angular position $\theta$ on the Earth's orbit
 and the date along the year is not correct in principle, as we will show in Sec.~\ref{Keplersections}. Earth does not revolve at constant rate during the complete orbit (that is the  Kepler's second law). However, we will show that $\theta$ given by Eq.~\eqref{thetaconstantrate}  is a very good approximation for Earth's orbit.

By using the dates of each photograph, in Table \ref{bosons2} we show that angular position of Earth along the trajectory for  different dates in which the photographs were taken.
\begin{table}[htp]
\begin{ruledtabular}
\begin{tabular}{c c c}
photo   & $t$  &  $\theta$  \\
\hline	
1& 83 &  1.428   \\
2&  125 &  2.150  \\
3& 138& 2.374\\
4&  157 & 2.701   \\
5&  170 &  2.924
\end{tabular}
\end{ruledtabular}
\centering
\caption{Angular position $\theta$ of Earth at different dates, given by Eq.~\eqref{thetaconstantrate}.}
\label{bosons2}
\end{table}

Now we have all the required data to calculate the form of the orbit of Earth.  We are able to obtain the five  distances $r$ and the angular positions $\theta$ extracted from Tables \ref{bosons} and \ref{bosons2} for the five photographs of the Sun
\begin{eqnarray}\label{netdata}
r_1&=& 149118984.1571\mbox{ [Km]} \, ,\quad \theta_1 = 1.428\, ,\nonumber\\
r_2&=& 150840671.5767\mbox{ [Km]} \, ,\quad \theta_2 = 2.150 \, ,\nonumber\\
r_3&=& 151364293.7829\mbox{ [Km]} \, ,\quad \theta_3 = 2.374 \, ,\nonumber\\
r_4&=& 151840565.2067\mbox{ [Km]} \, ,\quad \theta_4 = 2.701 \, ,\nonumber\\
r_5&=& 151950060.0148\mbox{ [Km]} \, ,\quad \theta_5 = 2.924 \, ,
\end{eqnarray}
where the subindex  of  distances and the angular positions 
is related to the number of each photograph.
These data will be useful in the next sections in order to find the correct form of the orbit. In Sec.~\ref{polarajuste},
we show that the data for the position obtained from the photographs describe a ellipse in polar coordinates. Also,  we show that the same elliptical orbit can be obtained in cartesian coordinates.  
This orbit is in very good agreement with the ellipse with correct values for eccentricity  and semimajor axis of the Earth's orbit \eqref{datossolnasaEx2} and \eqref{datossolnasasemiejer}.

\section{Orbit of the Earth from five photographs}
\label{polarajuste}

Using polar coordinates, we can develop an analytical process in which the equation for an ellipse is  shown to be the equation that fit the previous data \eqref{netdata}. This is achieved  obtaining the  ellipse parameters by an optimization scheme that resut to be in agreement with  values \eqref{datossolnasaEx2} and \eqref{datossolnasasemiejer}.

The equation of an ellipse, written in polar coordinates, is \cite{bate}
\begin{equation}\label{eqelipsegeneral}
r(\theta)=\frac{a \left(1-\epsilon^2\right)}{1+\epsilon\cos\theta}\, ,
\end{equation}
where $r$ and $\theta$ are depicted and defined in Fig.~\ref{coordinatesfigure}. Here, $a$ is the semimajor axis of the ellipse. The eccentricity is defined as
\begin{equation}
\epsilon=\sqrt{1-\frac{b^2}{a^2}}\, ,
\end{equation}
in terms of the semiminor axis  $b$  (see Fig.~\ref{coordinatesfigure}). This equation implies that the Sun is at one of the focus, and that the distance from Sun to Earth changes during the trajectory (as the angle changes). An eccentricity $\epsilon=0$ implies that the orbit is a circle, as $b=a$.

The purpose of this section is to show that the data points \eqref{netdata} are points in an ellipse \eqref{eqelipsegeneral}
for parameters $a$ and $\epsilon$ that are in agreement for those of Earth's orbit.
In order to prove this, we construct an optimization procedure that allow us to calculate the best fit of parameters $a$ and $\epsilon$ to the data \eqref{netdata}. Let us define the error function
\begin{equation}\label{errorfunction}
E(a,\epsilon)=\sum_{i=1}^5 \left(r_i -\frac{a \left(1-\epsilon^2\right)}{1+\epsilon\cos\theta_i} \right)^2\, ,
\end{equation}
where the sum is on $r_i$ and $\theta_i$ for $i=1,2,3,4,5$,
the five data points displayed in \eqref{netdata}. This error function measures how far  the data points are from fit an ellipse equation \eqref{eqelipsegeneral}. To have the best fit parameters, this function must be minimum. This is achieved by applying a minimization process to $E$ with respect to $a$ and $\epsilon$. Analytically, this implies that the partial derivatives of $E$ with respect to $a$ and to $\epsilon$ are both simultaneously null. The condition $\partial E/\partial a=0$ can be written as
\begin{equation}\label{condminimiza1}
\sum_{i=1}^5\frac{r_i}{1+\epsilon \cos\theta_i}=\sum_{i=1}^5 \frac{a\left(1-\epsilon^2\right)}{\left(1+\epsilon \cos\theta_i\right)^2}\, ,
\end{equation}
while the condition $\partial E/\partial \epsilon=0$ leads to 
\begin{eqnarray}\label{condminimiza2}
&&\sum_{i=1}^5\frac{r_i\left(2\epsilon+\left(1+\epsilon^2\right)\cos\theta_i\right)}{\left(1+\epsilon \cos\theta_i\right)^2}\nonumber\\
&&\qquad=\sum_{i=1}^5 \frac{a\left(1-\epsilon^2\right)\left(2\epsilon+\left(1+\epsilon^2\right)\cos\theta_i\right)}{\left(1+\epsilon \cos\theta_i\right)^3}\, .
\end{eqnarray}
Eqs.~\eqref{condminimiza1} and \eqref{condminimiza2} must be solved for $a$ and $\epsilon$. Those values minimize the error function $E(a,\epsilon)$ given in \eqref{errorfunction}. Performing both sumation  \eqref{condminimiza1} and \eqref{condminimiza2} on the five data points \eqref{netdata} in order to solve the above equations for $\epsilon$ and $a$ can be done by analytical methods or by using computational programs. Thereby, using data \eqref{netdata},
we obtain that the parameters that minimize the error function $E$
are
\begin{eqnarray}\label{datosellipeanalitical}
\epsilon&=& 0.0169\, ,\nonumber\\
a&=& 1.4952\times 10^8 \mbox{ [Km]}\, .
\end{eqnarray}
These values correspond to  eccentricity and the semimajor axis  of the ellipse described by Eq.~\eqref{eqelipsegeneral} that best fit the data points \eqref{netdata}.
Note the excellent agreement with the correct values \eqref{datossolnasaEx2} and \eqref{datossolnasasemiejer} for Earth's orbit. We can calculate the error of our prediction with respect to the  values \eqref{datossolnasaEx2} and \eqref{datossolnasasemiejer}. We find that for the eccentricity, the percentage error is about of $\left|({\epsilon_E-\epsilon})/{\epsilon_E}\right|\approx 1.15\%$. For the semimajor axis, the percentage error is about of $\left|({a_E-a})/{a_E}\right|\approx 0.05\%$
On the other hand, the error of our fit can be quantitative estimated by calculating the coefficient of determination
\begin{equation}
R^2=\left(\sum_{i=1}^5 \left(r(\theta_i)-\bar r\right)^2\right)\left(\sum_{i=1}^5 \left(r_i-\bar r\right)^2\right)^{-1}={0.9923}\, ,
\end{equation}
where $\bar r=(1/5)\sum_{i=1}^5 r_i$ is the average of our data, and $r(\theta_i)$ is the function \eqref{eqelipsegeneral} evaluated in the data points \eqref{netdata} with the found parameters \eqref{datosellipeanalitical}. The value for coefficient $R^2$ tells us that our fit is very good.

\begin{figure}[h!]
\centering
\includegraphics[width=8.5cm]{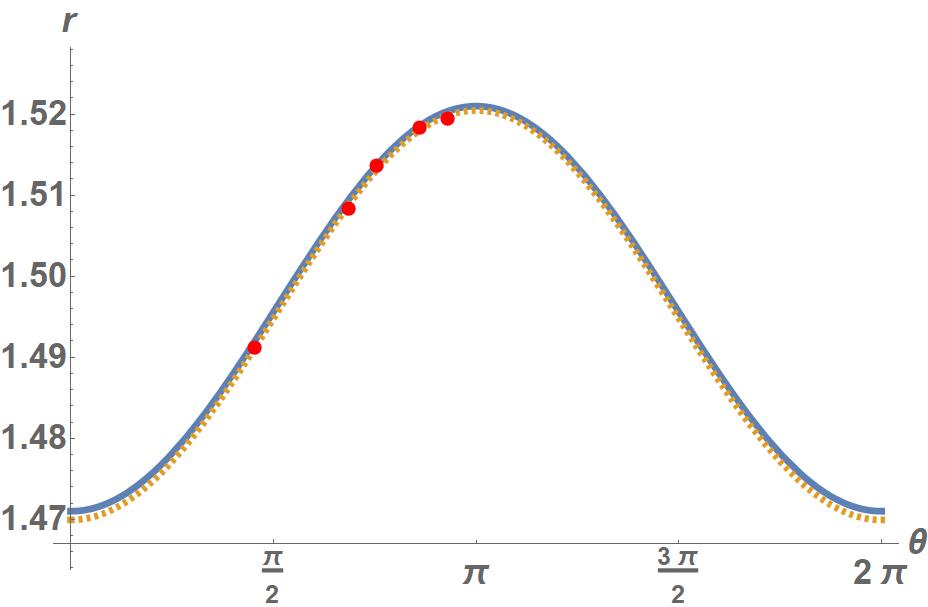}
\caption{The plot show the radii (measured in units of $10^8$[Km]) dependence with angle $\theta$ for an elliptical orbit in polar coordinates. The blue solid line is the known Earth's elliptical orbit \eqref{eqelipsegeneral}    with parameters \eqref{datossolnasaEx2} and \eqref{datossolnasasemiejer}. The yellow dashed  line is the  Earth's elliptical orbit \eqref{eqelipsegeneral} predicted for parameters  \eqref{datosellipeanalitical}.
The five data points \eqref{netdata} correspond to red circle points. Notice the striking fitting.}
\label{ajustelipsepountosanaliti}
\end{figure}

Results \eqref{datosellipeanalitical} imply that the five points \eqref{netdata} correspond to points in an ellipse. This is shown in Fig.~\ref{ajustelipsepountosanaliti} where the predicted elliptical orbit with parameters \eqref{datosellipeanalitical} (yellow dashed  line) is compared with the correct Earth's orbit (blue solid line). Notice how the data points  \eqref{netdata}, in red solid circles, fit very well in both curves.

\begin{figure}[h!]
\centering
\includegraphics[width=8cm]{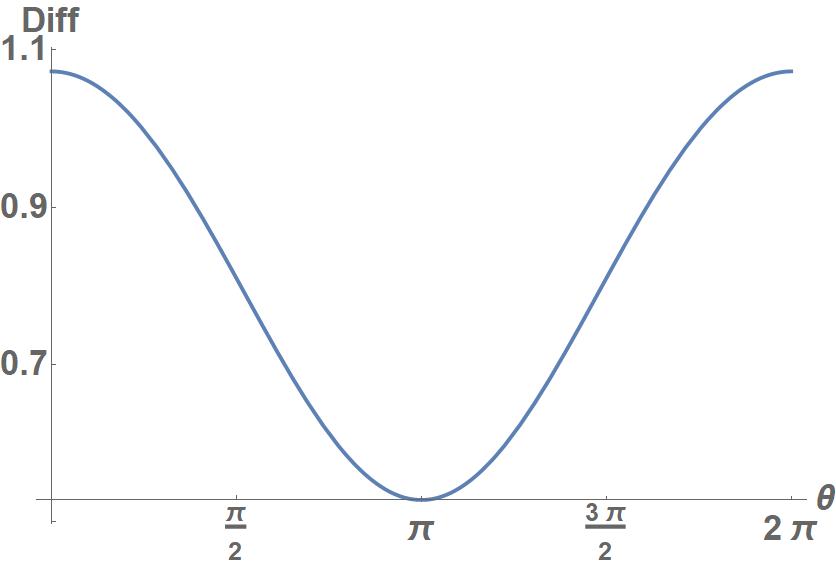}  
\caption{Plot of the difference function Diff given in Eq.~\eqref{diffducntion} in units of $10^5$[Km]. }
\label{errorelipsesanalit2}
\end{figure}

On the other hand, in order to evaluate our error in the determination the Earth's orbit, we can construct  a difference function between the correct orbit [with parameters  \eqref{datossolnasaEx2} and \eqref{datossolnasasemiejer}] and our calculation due to optimization of the five data points. This function reads
\begin{equation}\label{diffducntion}
\mbox{Diff}(\theta)=\left|\frac{a_E \left(1-\epsilon_E^2\right)}{1+\epsilon_E\cos\theta}-\frac{a \left(1-\epsilon^2\right)}{1+\epsilon\cos\theta}\right|\, .
\end{equation}
The absolute value of this difference will tell us information about where our estimation error is larger. The difference function $\mbox{Diff}$ is plotted in 
Fig.~\ref{errorelipsesanalit2}, from {where we see that the error margin is larger in $\theta=0$ and $2\pi$, i.e., at the perihelion. However, this error is about $\sim10^5$[Km], and thus our estimation has an approximated maximal percentage error of about
 $0.1$\%. On the other hand, near $\theta\sim\pi$, in the lapse of time where we took the photographs, the error decreases.}

\begin{figure}[h!]
\centering
\includegraphics[width=8.7cm]{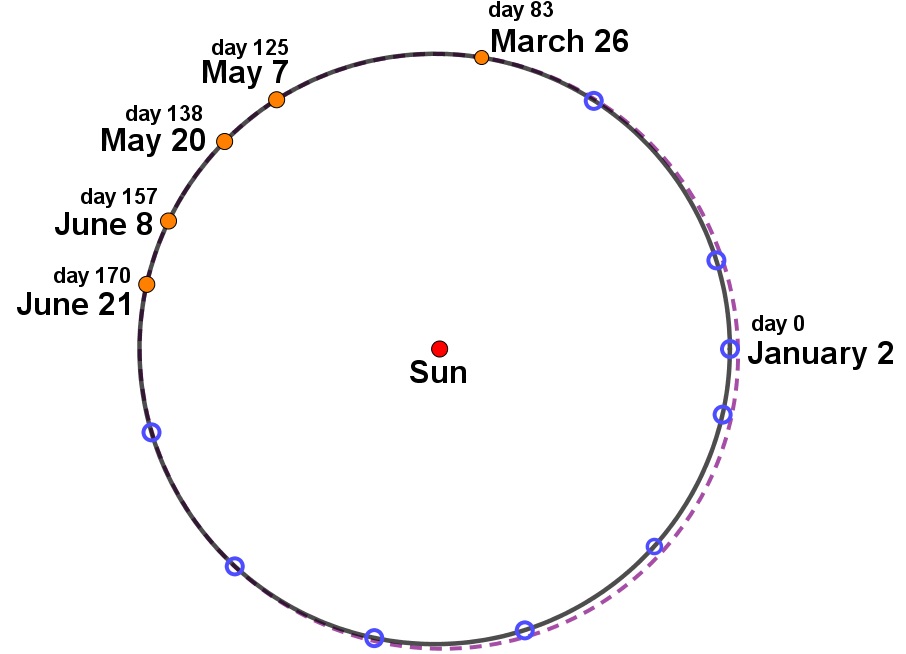}  
\caption{Earth's orbit is cartesian plane. Orange circle are the points \eqref{netdata}. Black solid  line is the predicted elliptical orbit. Blue hollow circles are future positions of Earth taken from  Stellarium software. Our predicted orbit fit all the points. The dashed purple line is the conic that fit the  points \eqref{netdata}, but that it does not coincide with the predicted elliptical orbit.}
\label{errorelipsesanalit5}
\end{figure}

In order to show that the predicted elliptical orbit of Earth coincide with the real one in a more appealing fashion, we plot the data points \eqref{netdata} in a cartesian plane using GeoGebra software \cite{geogebra}. This is shown in Fig.~\ref{errorelipsesanalit5}, where the data points \eqref{netdata} now represent the different position of Earth along the elliptical trajectory. The data points (in cartesian coordinates) are in orange solid circles, while the sun is at one of the focus of the orbit. The black solid line is the elliptical orbit predicted by parameters \eqref{datosellipeanalitical}, and as we can see the predicted orbit pass for all the five points.
Another interesting feature of Fig.~\ref{errorelipsesanalit5} is shown through in blue hollow  circles. These points correspond to different future Earth's positions extracted from  Stellarium software \cite{estela}, and they lie in the orbit fulfilled by parameters \eqref{datosellipeanalitical}. In this way, our predicted orbit  also foresee the future positions of Earth. Furthermore, in dashed purple line, the elliptical conic curve that fit the five data points \eqref{datosellipeanalitical} is shown. In principle, always a conic curve pass by five points. However, in our case, this conic curve does not fit the elliptical orbit predicted by  parameters \eqref{datosellipeanalitical}. The main reason is due to the proximity (in time) of our data. In order to use the treatment of finding an elliptical orbit by fitting a conic curve with five points, these data points should be obtained along a more extended span of time (a year), and not only approximatelly three months.

\section{Mass of the Sun and Kepler's laws}
\label{Keplersections}

In above section, we have proved  that the position of Earth (with its  distances and angles at several dates) correspond to points in an elliptical orbit where the Sun is in one focus. Our estimations of parameters of the elliptical orbit are in very good agreement with the real values for Earth's orbit. Therefore, we have proved the  Kepler's first law. This law establishes that elliptical orbits are direct consequence of the Newton central gravitational force (that is inverse to the square of the distance) owe to the central star  \cite{prentis,yasou}.

Even more interesting is the  Kepler's third law \eqref{secondKlaw}, that relates the form of the ellipse with the mass of the central star (in our case, the Sun).
By using Eq.~\eqref{secondKlaw}, we can calculate the Sun's mass. 
In Sec.~\ref{polarajuste} we already have found $a$ in Eq.~\eqref{datosellipeanalitical}. Using this, altogheter with the period $T$ given in \eqref{period}, we can calculate the Sun's mass. This calculation  is simply
\begin{equation}\label{masasun1}
{M}=\frac{4\pi^2 a^3}{G T^2}=1.9859\times 10^{30}\mbox{[Kg]}\, ,
\end{equation}
where $T$ have been used in seconds \eqref{period}. Compare this mass estimation with the real Sun's mass value given in 
\eqref{datossolnasaMASA}. This is the most important results of this work. With only five photographs, and using Kepler's laws,
we have been able to estimate the mass of Sun, with a percentage error of about $|(M_{\odot}-M)/M_{\odot}|\sim 0.13\%$. This shows that with only five photographs, the mass calculations are strikingly accurately.

Lastly, we are in position to discuss the  Kepler's second law. 
This law states that a line between the sun and the planet sweeps equal areas in equal times. This is a consequence of the conservation of the angular momentum of the planet. Basically, it establishes that any planet moves faster when is closer to the Sun, and slower when is far, implying that the angular velocity is not constant, i.e., a planet does not revolve around Sun a constant rate. In mathematical terms, the  Kepler's second law for Earth (conservation of its angular momentum) is written as $r^2 (d\theta/dt)=\sqrt{G\, a_E(1-\epsilon_E^2) M_\odot}$, where $d\theta/dt$ is the derivative of the angular position, i.e., how it changes in time \cite{bate}. Using the Kepler's laws for the elliptical form of the orbit, this equation can be put in the form
\begin{equation}
\frac{d\theta_E}{dt}=\frac{2\pi}{T}\frac{\left(1+\epsilon_E\cos\theta\right)^2}{\left( 1-\epsilon_E^2\right)^{3/2}}\, .
\end{equation}
Compare this result with Eq.~\eqref{thetaconstantrate} for an orbital motion at constant rate. Taking the derivative of Eq.~\eqref{thetaconstantrate}, we find
\begin{equation}
\frac{d\theta}{dt}=\frac{2\pi}{T}\, .
\end{equation}
This was our assumption of revolutions at constant rate
 in the calculations of Sec. III. Nevertheless, one can notice that because $\epsilon_E\ll 1$, the maximum error of our assumption (occurring in the perihelion) is of the order
$\left|({d\theta_E/dt-d\theta/dt})/({d\theta_E/dt})\right|\sim 2\epsilon_E$,
i.e., the maximum percentage error of considering Earth moving at constant rate is about of 3.4\%. As the Earth moves away from the perihelion, that error can be very small. For example, for photograph 1 at March 26th (day 83 and angle $\theta\approx1.4278$), the error is $\sim |2\epsilon_E\cos\theta|=0.48$\%. Therefore, our assumption \eqref{thetaconstantrate} for revolutions at constant rate is justified, and in agreement with the  Kepler's second law for percentage errors of the order of 1\%. This is the reason why our results are so good even when strictly we are violating the  Kepler's second law.

\section{Conclusions}

With this work, we have shown that in a lapse of time of about three months, several features about the Earth's trajectory and Sun can be obtained from only five photographs. Any student can perform the present analysis under the guidance of a teacher. Students perfoming this experience will
realize that  observations from Earth allow to calculate distances to other bodies. 
Most important, this experience combine practical astronomical observations and theoretical physics knowledge with analytical and computational skills. These are the kind of proficiencies that any student should possess.

\begin{figure}[h!]
\centering
\includegraphics[width=7cm]{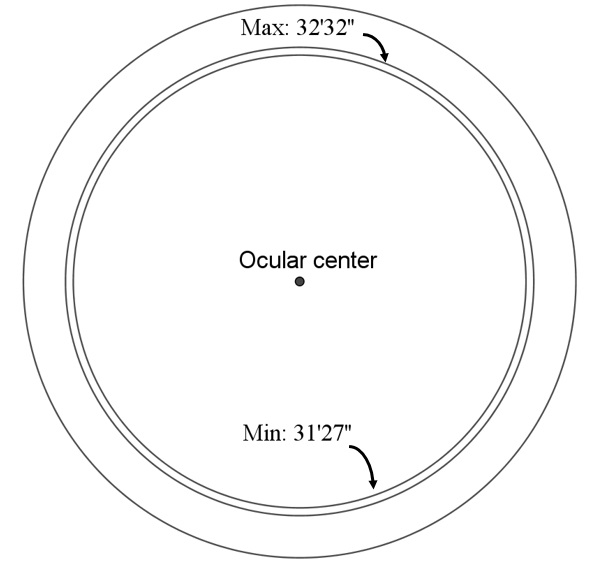}  
\caption{Maximum and minimum of Sun's angular sizes}
\label{minmax}
\end{figure}

Several following comments are useful to anyone attempting to repeat this experience. First, with the five point presented here, we were able to calculate the elliptical orbit \eqref{eqelipsegeneral} that assumes that the Sun is at a focus of the ellipse. With more photographs, or the same amount of photographs taken in a large lapse of time, it is even possible to prove that the Sun is at one focus of the ellipse. Secondly, we do not recommend to take data points in a small lapse of time, as that increases the magnitude of the error estimation.
Also, the photographs should be taken in a sunny day with the Sun  in its highest position, as this decreases the  atmospheric aberration. On the other hand, in general, all the photographs in Fig.~\ref{eclipseFig} have errors associated to the measurements of $r_o$ and $r_{aS}$. One can improve the accuracy in their calculation, by counting the pixels
in the photographs. This allow us to minimize the error associated to the focus of photographs. This is important as the Sun (when is inside the telescope ocular) varies its angular size very little during a year. In Fig.~\ref{minmax} is shown the   Sun's  maximum $32'32''$ and minimum $31'27''$ angular sizes. Therefore, the difference between Sun's maximum and minimum angular sizes is $1'5''$ or $\ang{0.0181}$. For this experience is imperative
a correct calculation of $r_{aS}$ through a  careful measuments of the angles.

Lastly, it is important the quality of the telescope solar filter in order to have good photographs. Also, a better focusing of the photographs can be obtained when solar spots are present.

On the other hand, the eccentricity can be measured by only knowing the position of Earth's perihelion and aphelion. By taking photographs in those points, geometrical method can be used to estimate the eccentricity. However, for this approach, the ellipse orbit is assumed, and those two points do not provide enough information to calculate the orbit's parameters. On the contrary,
following the same procedure of this paper, the orbits of Moon and Jupiter can be calculated, as these bodies are visible from Earth using telescopes. These works are left for the future.

Finally, it is also important to consider that Sun's diameter $D_\odot$ is considered a known variable along this work. Its value  can be estimated experimentally, although its precision depends on several solar factors \cite{sund}, which implies that it needs to be estimated periodically. However, a simple and very accurate manner to estimate Sun's diameter can be constructed by using an eclipse. The ratio between Sun and Earth radii can be estimated using the tecniques developed in Ref.~\cite{caerolsasenjomooon}. Thus, by knowing Earth's diameter (for example, through Eratosthenes' experiment), the Sun's diameter is readily calculated.

All the above considerations lead to or improve the results shown along this work. 
We strongly believe that this experience can be developed in a simpler manner by a group of students.
  The process of taking the photographs, the use and understanding of a telescope,
the computational analysis to measure the radii, the analytical methods to evaluate the orbit, and the theoretical use of Kepler's laws to estimate the Sun's mass can be used by a teacher to encourage  the scientific knowledge pursue in students, and at the same time, to explain to them the different processes that a scientist must do in order to unveil deep truths behind the facts.

\begin{acknowledgments}
 F.A.A. was supported by Fondecyt-Chile Grant No. 1180139.
\end{acknowledgments}

\end{document}